\documentclass[%
reprint,
 amsmath,amssymb, 
 aps,prb,
]{revtex4-1}

\usepackage{mathtools} 
\usepackage{multirow} 
\usepackage{color} 
\usepackage{epsf} 
\usepackage{soul}
\usepackage{epsfig} 
\usepackage{ulem} 
\usepackage{graphicx}
\usepackage{dcolumn}
\usepackage{bm}
 
\begin{document}

\title{Half-metallic ferromagnetism in Sr$_3$Ru$_2$O$_7$}

\author{Pablo Rivero$^1$, Vincent Meunier$^2$, and William Shelton$^{1}$} 
 
\email{jprivero@lsu.edu}

\affiliation{1. Center for Computation and Technology, Louisiana State University, Baton Rouge, Louisiana 70803, USA \\ 
2. Department of Physics, Applied Physics, and Astronomy, Rensselaer Polytechnic Institute, Troy, NY 12180, USA }

\date{\today}

\begin{abstract}

\noindent{}The bilayered member of the Ruddesden-Popper family of ruthenates, Sr$_3$Ru$_2$O$_7$, has received an increasing attention due to its interesting properties and phases. By using first principle calculations we find that the ground-state is characterized by a ferromagnetic (FM) half-metallic state. This state strongly competes with an antiferromagnetic metallic phase, which indicates the possible presence of a particular state characterized by the existence of different magnetic domains. To drive the system towards a phase transition we studied the electronic and magnetic properties as a function of RuO$_6$ octahedra rotations and found that the magnetic phase does not couple with the rotation angle. Our results provide accurate electronic, structure, and magnetic ground-state properties of Sr$_3$Ru$_2$O$_7$ and stimulate the investigation of other type of octahedra rotations and distortions in the search of phase transitions.

\begin{description} 
\item[PACS numbers] 

 
\end{description} 
\end{abstract} 
 
\pacs{Valid PACS appear here}

\maketitle 
 

\section{I. Introduction} 

The Ruddesden-Popper (RP) Sr$_{n+1}$Ru$_n$O$_{3n+1}$ family of layered perovskites (where $n$ refers to the number of RuO$_6$ octahedra layers in the system \cite{RP}) have received an increasing amount of attention in the last two decades due to a number of interesting phases and properties. For instance, the bilayer member Sr$_3$Ru$_2$O$_7$ ($n$=2) shows metamagnetic transitions, \cite{metamagnetism1} quantum critical phenomena, \cite{metamagnetism2} and electronic nematic phases. \cite{nematic} In addition, the physical properties of the system can be manipulated via external parameters (temperature, pressure, magnetic field, ...) \cite{uniaxial,magneticfield,temperature} or by defects and alloying \cite{Ti,Ti2,Mn,partial1,partial2,partial3} to produce metal-to-insulator or magnetic phase transitions. To determine and control these properties requires an understanding of the ground-state structure and the role of RuO$_6$ octahedra rotations in the system.

Neutron powder diffraction measurements on Sr$_3$Ru$_2$O$_7$ have determined the existence of two symmetry related different space groups, $Pban$ \cite{pban} (\#50) and $Bbcb$ \cite{bbcb1,bbcb2} (\#68). The $Bbcb$ spacegroup is a subgroup of $Pban$. While $Bbcb$ has shown to yield a better fit with the scattering data, the $Pban$ structure allows for a larger number of AFM configurations. The Sr$_3$Ru$_2$O$_7$ structure is formed by two layers of RuO$_6$ octahedra connected by sharing an apical oxygen and interleaved by two SrO layers (see Fig. 1). The octahedra layers in the bilayer are both rotated about the $c$-axis with one octahedron rotated clockwise while the other is rotated counter-clockwise. 

Experimental measurements at room temperature show that each RuO$_6$ octahedron is rotated by 7.18$^{\circ}$. \cite{pban} As the temperature is lowered to 9K the rotation angle increases by 15$\%$ to 8.05$^{\circ}$ along with a small 1$\%$ reduction in the in-plane lattice parameters and a corresponding 1$\%$ increase in the $c$ parameter leading to a small reduction in volume. \cite{pban} This result indicates that octahedral rotations are sensitive to temperature and possibly to other thermodynamic parameters (i.e., pressure, composition, etc.). However, there are no observed structural or magnetic phase transitions within this temperature range and, unlike the surface Sr$_3$Ru$_2$O$_7$ structure, no octahedral tilts have been observed. \cite{ChenChen} 

At low temperature, Sr$_3$Ru$_2$O$_7$ is a paramagnetic (PM) metal with Fermi liquid state behavior that is close to a FM instability. \cite{paramagnet1,paramagnet2} In fact, it has been shown that a modest application of uniaxial pressure of about 0.1 GPa along the $c$-axis drives the system into a FM state. \cite{uniaxial} Moreover, by applying an external magnetic field in the range of 5 to 8 T parallel to the $ab$ plane gives rise to a rapid increase in the magnetization of the system (metamagnetism). \cite{magneticfield} In contrast, when Sr$_3$Ru$_2$O$_7$ is under hydrostatic pressure, the PM phase is stabilized. \cite{hydrostatic1,hydrostatic2} Furthermore, magnetic susceptibility measurements indicate a short-range AFM-type correlation at temperature below $\simeq$ 20 K. \cite{pban,afm} These results suggest the possibility of having two competing magnetic interactions (AFM and FM) at low temperature which, as we will show later, is in agreement with our calculations.

It is noteworthy to point out that the single and bilayer members ($n$=1 and $n$=2) of the ruthenate RP family are PM metals but larger layered members ($n$ \textgreater 2) are FM metals with localized magnetic moments ranging from 0.2 $\mu$B \cite{4310} to 1.6 $\mu$B. \cite{magnetic-SRO} Similarly, the conductivity at low temperatures is mainly confined in the $ab$ plane for the $n$=1 and $n$=2 members (PM systems) with a resistivity ratio ($\rho$$_z$/$\rho$$_{xy}$) of about 1000 \cite{ratio-Sr2RuO4} and around 300\cite{ferro1} respectively, which denote the quasi two-dimensional character in these early members of the family. In a comparison, Sr$_4$Ru$_3$O$_{10}$ (n=3) has a resistivity ratio of 31\cite{ratio4310} while SrRuO$_3$ ($n$=$\infty$) has an almost isotropic conductivity with a value of about 1.1.\cite{ratio-SrRuO3} These results indicate that paramagnetism occurs when the resistance along the $c$-axis increases, where the $n$=2 member is on the verge of a magnetic transition. This is important since presumably the application of certain external parameters could drive this system towards electronic and magnetic phase transitions. Understanding and controlling these transitions could have a significant impact on technological applications (e.g., transistors and memories with higher energy efficiencies).

\begin{figure}[t]
\includegraphics[width=0.45\textwidth]{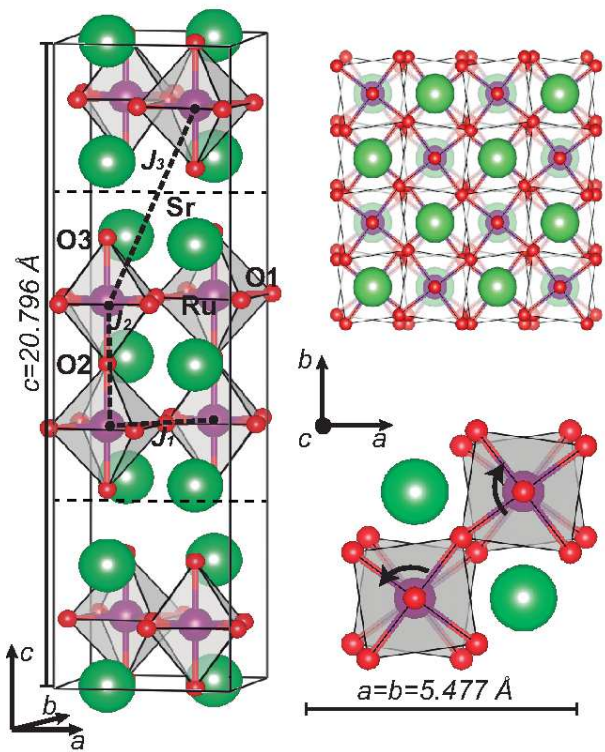}
\caption{Side and top views of the orthorhombic $Pban$ structure of Sr$_3$Ru$_2$O$_7$. The RuO$_6$ octahedra are rotated alternately clockwise and counter-clockwise about the $c$-axis. The structure does not exhibit any other axis rotation. Dashed lines delimit the Sr$_3$Ru$_2$O$_7$ bilayer. $J$$_1$, $J$$_2$, and $J$$_3$ are the magnetic couplings studied in this work. O1 refers to the in-plane oxygens, while O2 and O3 are the two different apical oxygens in the system. The experimental cell parameters displayed in the figure have been extracted from Ref. \onlinecite{pban}.}
\label{fig1}
\end{figure}

Previous computational investigations only considered the non-magnetic (NM) or FM phases for understanding the electronic and magnetic ground-state properties.\cite{singh,bosson1,bosson2,hetero} However, a more exhaustive study requires the investigation of a larger number of magnetic phases. In the present work we study the NM, the FM, and various AFM phases in Sr$_3$Ru$_2$O$_7$ to understand the energetics and the mechanisms that give rise to the observed properties of this system. This allows us to reveal a FM half-metallic ground-state structure. We also investigate the effects of RuO$_6$ octahedra rotations on the electronic and magnetic properties and find no magnetic phase transition independently of the rotation performed. These results lay the foundation for future studies on the effects of doping or pressure on the properties of bulk and surface structures, where additional rotational degrees of freedom (i.e., tilts) are allowed and may have a significant impact on the properties of the material.

\section{COMPUTATIONAL DETAILS}

All calculations were performed with the replicated-data version of the CRYSTAL14 computational package. \cite{CRYSTAL14,CRYSTAL142} CRYSTAL14 employs atom-centered Gaussian type orbital (GTO) basis sets to build Bloch functions, which are used to expand the one-electron crystalline orbitals. Atom centered basis that use GTOs can provide a significant improvement in computational cost when using hybrid functionals due to the natural cut-off with distance inherent in GTOs, as compared to plane-waves, which are global basis. 

The GTO basis sets for each atom comprising the Sr$_3$Ru$_2$O$_7$ system were taken from Ref. \onlinecite{webpage}. For Ru and Sr, the small-core Hay–Wadt pseudopotentials \cite{Hay} were adopted for the description of the inner-shell electrons \textit{(1s$^2$2s$^2$2p$^6$3s$^2$3p$^6$3d$^{10}$)}. The valence functions for Ru were based on the modified LANL2DZ basis: \cite{lanl2dz} \textit{4s$^2$4p$^6$4d$^7$5s$^1$},  while the \textit{4s$^2$4p$^6$5s$^2$} was used for Sr. Finally, for O atoms we used the 8-411$d$ all-electron basis set constructed by Cor\`a. \cite{cora} 

\begin{table*}[]
\centering
\caption{Cell parameters (\AA), rotation ($^{\circ}$), total energy differences: $\Delta$E$_{AFM-I}$ = E$_{AFM-I}$ - E$_{FM}$; $\Delta$E$_{AFM-A}$ = E$_{AFM-A}$ - E$_{FM}$; and $\Delta$E$_{NM}$ = E$_{AFM-NM}$ - E$_{FM}$  (meV), magnetic moment ($\mu$B), and volume (\AA$^3$) obtained via LDA, PBE, PBESol (PBES), PBE-10, and PBESol-HFX (PBES-HFX) hybrid functionals in comparison to experimental data (EXP) extracted from Ref. \onlinecite{pban}. Ru-O1 are the in-plane Ru-O distances while Ru-O2 and Ru-O3 are the correspondent to the out-of-plane (see Fig. 1). }
\label{my-label}
\begin{tabular}{lcccccccccccc}
                & a     & b     & c      & Rotation & Ru-O1    & Ru-O2  & Ru-O3 & magn. & $\Delta$E$_{AFM-I}$ & $\Delta$E$_{AFM-A}$ & $\Delta$E$_{NM}$ & Vol. \\ \hline
LDA             & 5.399 & 5.399 & 20.732 & 10.84    & 1.943        & 2.034  & 2.043 & 0.77  & 125  & 34.3 & 230  & 604.3 \\  
PBE             & 5.575 & 5.576 & 20.913 & 10.45    & 2.005        & 2.048  & 2.057 & 1.41  & 13.0 & 267  & 995  & 650.1 \\
PBE10           & 5.543 & 5.552 & 20.816 &  9.87    & 1.999,1.983  & 2.033  & 2.046 & 1.41  & 5.6  & 420  & 3490 & 640.6 \\
PBES          & 5.503 & 5.503 & 20.656 & 10.23    & 1.977        & 2.024  & 2.035 & 1.32  & 15.2 & 192  & 586  & 625.5 \\
PBES-5        & 5.498 & 5.499 & 20.613 & 9.98     & 1.975,1.973  & 2.016  & 2.029 & 1.385 & 16.6 & 345  & 1797 & 623.2 \\
PBES-10       & 5.494 & 5.486 & 20.594 & 9.71     & 1.964,1.975  & 2.010  & 2.025 & 1.39  & 2.1  & 402  & 3000 & 620.8 \\
PBES-15       & 5.485 & 5.477 & 20.573 & 9.54     & 1.955,1.975  & 2.006  & 2.022 & 1.394 & 6.6  & 462  & 4320 & 618.1 \\
PBES-20       & 5.475 & 5.464 & 20.566 & 9.39     & 1.947,1.975  & 2.005  & 2.019 & 1.41  & 4.1  & -    & 5630 & 615.3 \\ \hline
EXP             & 5.477 & 5.477 & 20.796 & 8.05     & 1.956        & 2.026  & 2.038 & -     & -    & -    & -    & 623.8 \\ \hline
\end{tabular}
\end{table*}

For our calculations we employed an 8 x 8 x 8 Monkhorst-Pack mesh \cite{monkhorst} that corresponds to 125 k-points in the irreducible Brillouin zone. The thresholds controlling the accuracy in the evaluation of Coulomb and exchange integrals were set to 10$^{-7}$ (ITOL1, ITOL2, ITOL3, and ITOL4, using notations from Ref. \onlinecite{CRYSTAL142}) and 10$^{-14}$ (ITOL5), while the SCF energy threshold was set to 10$^{-6}$ au. Geometry optimizations have been performed with the same 0.0003 Hartree/Bohr convergence criterion on gradient components and 0.0012 Bohr nuclear displacements. By using these parameters we obtain converged total energies within 1-2 meV per unit cell.

Hybrid functionals can provide a better description of the electronic exchange in materials where strong magnetic phase competition plays an important role on the electronic and magnetic structure and energetics. Therefore, besides some standard DFT functionals such as LDA (VWN \cite{VWN}), PBE \cite{pbe}, and PBEsol \cite{PBESOL}, we investigate the use of some hybrid functionals based on PBE and PBESol. For convenience the PBESol hybrid functionals will be written as PBES-5, PBES-10, PBES-15, and PBES-20, where the number indicates the percentage of HFX mixing used in the hybrid functional. 

\section{Results and Discussion}

\subsection{Electronic and magnetic properties}

At low temperature Sr$_3$Ru$_2$O$_7$ exhibits a PM metallic character. However, the theoretical approach used in this work is not capable of directly treating a PM state. This is due to the fact that electronic spins in magnetic calculations using CRYSTAL14 must be constrained to a defined orientation. Therefore, we study the FM, the NM, and 4 different AFM states (Fig. 2). The AFM states include: AFM-I (ferromagnetic bilayers coupled antiferromagnetically), AFM-A (ferromagnetic coupling in-plane coupled antiferromagnetically out-of-plane for each bilayer), AFM-C (antiferromagnetic coupling in-plane and ferromagnetic out-of-plane for each bilayer), and AFM-G (antiferromagnetic coupling in-plane and out-of-plane for each bilayer). This approach allows us to estimate the energetic stability and total energy difference of various magnetic phases to understand the electronic and magnetic structure of the system.

\begin{figure}[t]
\includegraphics[width=0.48\textwidth]{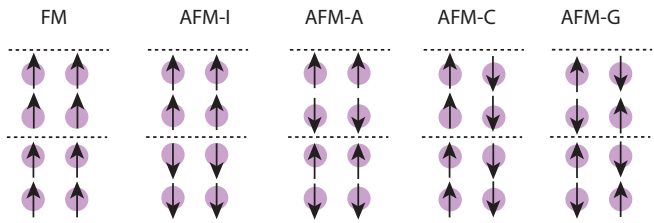}
\caption{Schematic of the magnetic orderings considered in our calculations. Dashed lines separate adjacent bilayers. Only Ru atoms are displayed.}
\label{fig2}
\end{figure}

In Table I we display the calculated structural parameters, magnetic moments, and total energy differences between FM, NM, and various AFM states using different DFT functionals. The energy differences are all relative to the FM state and all results were obtained after full geometry relaxation. Total energy differences that involve AFM-C and AFM-G phases ($\Delta$E$_{AFM-C}$ and $\Delta$E$_{AFM-G}$) have not been included in the table due to their much higher energies compared to the other AFM phases. All tested functionals yield a FM solution as the ground-state with a magnetic moment of $\simeq$ 1.4  $\mu$B except for LDA which predicts a significantly smaller moment of $\simeq$ 0.77 $\mu$B per Ru atom (in agreement with Ref. \onlinecite{singh}) and a smaller volume indicating overbinding, which is a common LDA problem in solids.

\begin{figure}[t]
\includegraphics[width=0.35\textwidth]{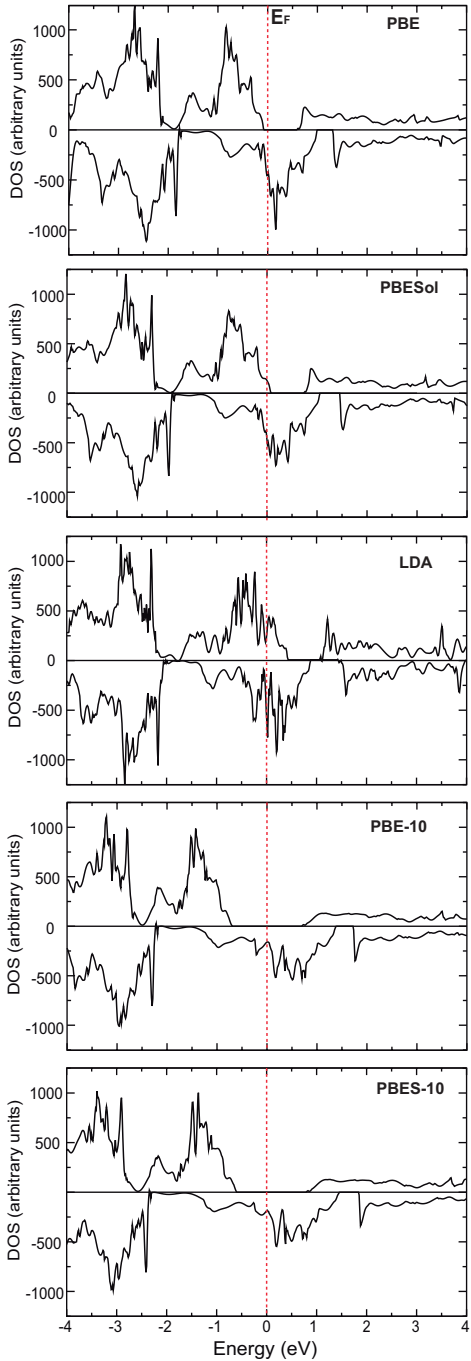}
\caption{Total density of states obtained via PBE, PBESol, LDA, PBE-10, and PBES-10 functionals. While PBE, PBE-10, and PBES-10 produce a half-metal, LDA and PBESol yield a metal.}
\label{fig3}
\end{figure}

Interestingly, we find that the FM state is nearly-degenerate with the AFM-I phase ($\Delta$E$_{AFM-I}$ $\simeq$ 2-16 meV). Both phases practically have the same structural parameters with relative differences less than 0.2$\%$. To ensure that the small energy difference is reliable, we performed additional simulations using a 16 x 16 x 16 k-space grid along with increasing tolerances to 10$^{-7}$ au on the full geometry relaxation and observed minor changes of the order of 1 meV. This result highlight the fact that at low temperature there is a strong competition between AFM-I and FM phases indicating the possibility of having a mixed state characterized by different magnetic domains.

\begin{figure}[t]
\includegraphics[width=0.45\textwidth]{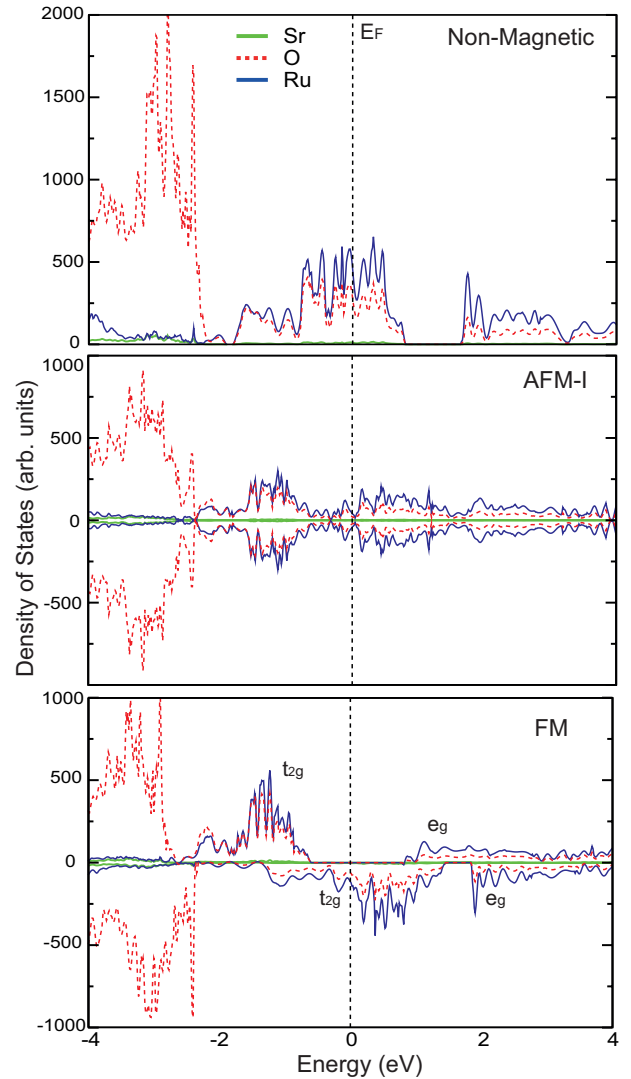}
\caption{Projected density of states of the fully relaxed NM, AFM, and FM orthorhombic phases using PBES-10 functional.}
\label{fig4}
\end{figure}

All functionals used in this study provide accurate structural parameters as compared with experiment. With PBESol, these parameters are slightly closer to experiment than those obtained with PBE, improving the average relative error from $\varepsilon$$_{rel.}$ = 5.5$\%$ to 4.0$\%$. We also note that a small amount of HFX added to the PBESol further reduces the cell parameters, volume, and rotations bringing the results into better alignment with experiment with a $\varepsilon$$_{rel.}$ = 2.7$\%$ (PBES-10). Curiously, the density of states (DOS) obtained with these functionals is different. PBESol produces a metal while PBE (and PBES-10) predicts a half-metal (Fig. 2). We find the same results when performing full geometry relaxations via PBE and PBESol using the Vienna Ab-Initio Simulation Package (VASP\cite{VASP}). This is a consequence of the different exchange enhancement factors ($F$$_x$) fitted in both functionals. $F$$_x$ is used to scale the gradient dependence of the exchange energy of the local density approximation. 
In PBESol the $F$$_x$ was fitted to improve the lattice constants of solids over PBE, but degrades the atomization energies of molecules and violates the asymptotic expansion of the exchange energy of neutral atoms.\cite{PBESOL} PBESol has a weaker $F$$_x$ value than PBE and therefore, has an exchange energy approximation closer to LDA's. As a result PBESol produces electronic structure properties that are in between PBE and LDA as it can be seen in Fig. 3 for the DOS. When a nonlocal Hartree-Fock exchange (HFX) term is introduced to the PBESol functional the half-metallic character is recovered.

In general, both PBE or PBES-HFX provide similar performance in the structural, electronic, and magnetic properties. However, since our calculations will be focused on doped Sr$_3$Ru$_2$O$_7$ systems (e.g., Mn, etc) and the study of phase transitions, hybrid functional calculations may be required. Moreover, preliminary calculations on the surface Sr$_3$Ru$_2$O$_7$ structure \cite{our-surface} show that, unlike PBE, PBES-10 captures the RuO$_6$ octahedra tilts at the surface in agreement with experiment measurements. \cite{ChenChen} Therefore, in order to make it possible to compare with our future investigations we will employ PBES-10 for the reminder of the paper.

Let us now study the electronic structure of Sr$_3$Ru$_2$O$_7$. In Figure 4 we show the projected density of states (PDOS) for the NM, FM, and AFM phases obtained with PBES-10. The NM PDOS result is in perfect agreement with previous theoretical works.\cite{singh,bosson1,bosson2} However, according to our calculations, the FM state is the ground-state structure by a total energy difference of 3 eV. This state features a half-metallic character with a spin $\alpha$ bandgap of about 1.5 eV. The half-metallic character observed in this system is consistent with other investigations on different perovskite systems.\cite{our1,our2,other1} Finally, the metallic state of the AFM-I configuration is derived from the mixture of electronic states that come from neighboring bilayers (i.e., each bilayer has a half-metallic character with alternating spins). Singh et al. proposed this AFM state (our AFM-I) as the ground-state structure for Sr$_3$Ru$_2$O$_7$ based on the NM and FM simulations and experiment measurements.\cite{singh} However, our hybrid DFT functional calculations predict an FM ground-state solution that is 2 meV (that corresponds to $\simeq$ 22K) lower in energy than the AFM-I state (Table I). 

Hitherto, we have found that Sr$_3$Ru$_2$O$_7$ is characterized by a half-metallic state with a strong competition between FM and AFM-I phases. Therefore, a better understanding of the strength of the magnetic interactions occurring in the system is essential to control phase transitions.
To determine the coupling strength between different neighboring Ru atoms we extract some magnetic coupling parameters from the FM and AFM energies. \cite{magnetic-coupling} These parameters can be obtained by means of the Ising Hamiltonian. \cite{Ising} This Hamiltonian, that emerges as a simplification of the Heisenberg Hamiltonian,\cite{Heisenberg} only commutes with the $z$-component of the total spin operator and is expressed as,

\begin{equation}
\hat{H}_{Ising}=-\sum_{ij} J_{ij} \hat{S}_{zi} \hat{S}_{zj} 
\end{equation}

$\noindent{}$where $\Sigma$ $ij$ denotes a sum over all equivalent pairs of first, second, etc. neighbor sites and \textit{J$_{ij}$} is the exchange coupling constant between $\hat{S}$$_{zi}$ and $\hat{S}$$_{zj}$ localized spin moments. According to Eq. 1 a positive value of \textit{J$_{ij}$} corresponds to a FM interaction while a negative sign indicates a AFM interaction. 

Here, we consider three relevant magnetic couplings: \textit{J$_1$}, \textit{J$_2$}, and \textit{J$_3$}. \textit{J$_1$} is the coupling between Ru nearest neighbors, which are localized in the same plane; \textit{J$_2$} is the coupling between next nearest neighbors, localized out-of-plane in the bilayer; and \textit{J$_3$} is the coupling corresponding to the nearest Ru atoms localized in different bilayers (see Fig. 1). The magnetic coupling parameters are calculated by mapping the energy differences of different spin arrangements to the Ising Hamiltonian (Eq. 1). The 48 atom Sr$_3$Ru$_2$O$_7$ ($Pban$) unit cell, where each Ru atom has two unpaired electrons per Ru center (S$_z$=1), leads to the following equations:

\begin{equation}
J_1= - \frac{1}{40} [ E_{FM}-E_{AFM-C} ]
\end{equation}
\begin{equation}
J_2 = - \frac{1}{8} [E_{AFM-I}-E_{AFM-A} ]  
\end{equation}
\begin{equation}
J_3 = - \frac{1}{20} [E_{FM}-E_{AFM-I} ]
\end{equation}

Solving these equations using the calculated energies we obtain \textit{J$_1$}= +30.1 meV, \textit{J$_2$}= +51.1 meV, and \textit{J$_3$}= +0.19 meV (per pair of Ru atoms). These values indicate: 1- All three magnetic couplings are ferromagnetic, 2- The coupling between bilayers is weak and suggests the possible coexistence of FM and AFM-I phases, and 3- the FM out-of-plane coupling (\textit{J$_2$}) is the strongest magnetic interaction. This can be understood as a consequence of the stronger polarization between Ru d$_{xz}$ and d$_{yz}$ orbitals with the O p$_z$ through the 180$^{\circ}$ Ru-O-Ru along the $c$-axis in comparison to the Ru-d$_{xy}$ and O p$_x$, p$_y$ orbitals forming 160$^{\circ}$ in-plane.

\begin{figure}[t]
\includegraphics[width=0.45\textwidth]{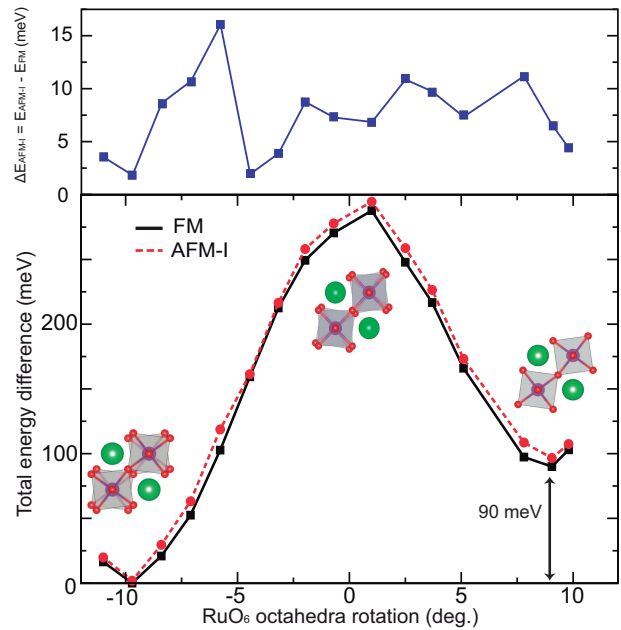}
\caption{Evolution of the total energy difference between the FM (and AFM-I) phase and the ground-state structure as a function of the RuO$_6$ octahedron rotations only in the top layer of the bilayer (\textit{partial rotation} case).}
\label{fig1}
\end{figure}

Based on the Goodenough-Kanamori rules, the Ru$^{4+}$-Ru$^{4+}$ coupling should be AFM. \cite{GK1,GK2} However, Sr$_3$Ru$_2$O$_7$ has itinerant carriers and magnetic moments, and therefore, a non-negligible double-exchange (DE) FM interaction. This interaction is mediated via Hund's rule coupling between itinerant electrons and localized moments. \cite{DE} Based on these results the FM coupling is dominant, which favors the DE interaction over superexchange which mainly applies to localized states. Reducing the DE interaction by, for example, partial substitution of Ru by Ti --atomic center acting as a nonmagnetic impurity with similar ionic radius-- favors the localization of itinerant electrons modifying the structure of the system and giving rise to an AFM state. \cite{Ti} Therefore, octahedra rotations and distortions could also trigger a phase transition.

\begin{figure*}[t]
\includegraphics[width=0.90\textwidth]{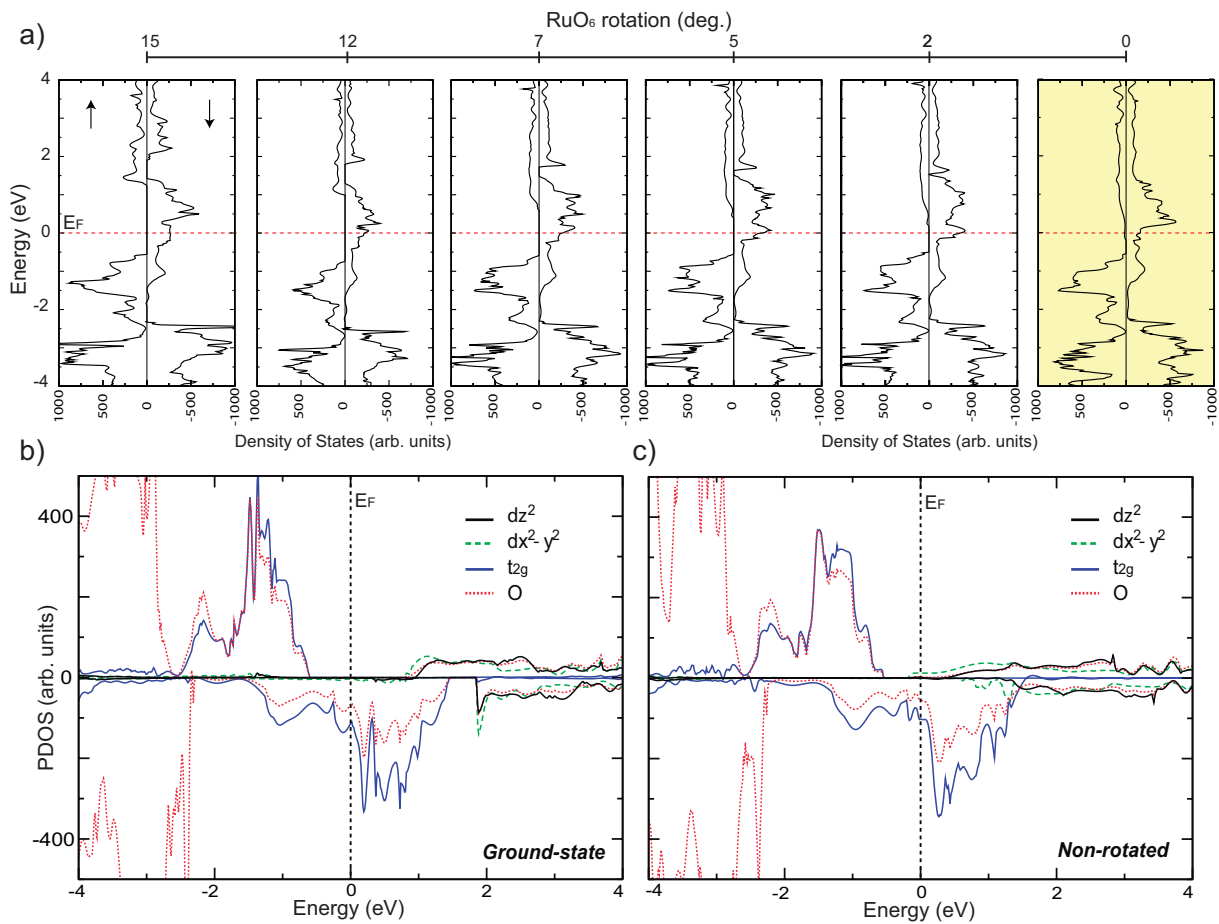}
\caption{(a) Evolution of the DOS for the FM state as the RuO$_6$ octahedra layers rotate (\textit{simultaneous rotation} case). The half-metallic character disappears only for the system without rotations. (b) Projected density of states onto O and Ru-4$d$ orbitals for the calculated ground-state structure (9.7$\circ$) and (c) the hypothetical non-rotated Sr$_3$Ru$_2$O$_7$ structure.}
\label{fig6}
\end{figure*}

\subsection{Role of RuO$_6$ rotations}

In this section we investigate the potential of obtaining an energetically favorable AFM structure by breaking the t$_{2g}$ and e$_g$ symmetries via octahedron rotations. We consider two types of rotations: where we systematically vary the rotation angle of one of the RuO$_6$ octahedron layers in the bilayer while keeping the other fixed (\textit{partial rotation}), and where both octahedron layers in the bilayer are rotated in the opposite direction and by the same amount (\textit{simultaneous rotation}). In this study we performed atomic and structural relaxations as a function of the RuO$_6$ octahedron angles.

\begin{table}[]
\centering
\caption{Structural parameters (\AA), total energy difference (meV), magnetism ($\mu$B), and volume (\AA$^3$) as a function of the RuO$_6$ octhedron rotation angles obtained via PBES-10 (\textit{simultaneous rotation} case). }
\label{my-label}
\begin{tabular}{ccccccc}
\textbf{Rotation} & \textbf{$\Delta$E$_{AFM-I}$} & \textbf{a} & \textbf{b} & \textbf{c} & \textbf{vol} & \textbf{magn. }  \\ \hline
0.0          & 2               & 5.564      & 5.563      & 20.104 & 622.3    & 1.36        \\
2.1          & 5                & 5.561      & 5.562      & 20.109 & 622.0    & 1.36        \\
4.5          & 6             & 5.537      & 5.545      & 20.258 & 622.0    & 1.37        \\
7.1          & 10              & 5.526      & 5.536      & 20.337 & 622.1    & 1.37        \\
9.7          & 2                & 5.494      & 5.486      & 20.594 & 620.7    & 1.39        \\
11.7         & 5                & 5.466      & 5.459      & 20.764 & 619.6    & 1.40        \\
12.3         & 11             & 5.451      & 5.448      & 20.840 & 618.9    & 1.40        \\
13.1         & 9               & 5.431      & 5.433      & 20.942 & 617.9    & 1.41        \\
14.1         & 11              & 5.409      & 5.417      & 21.055 & 616.9    & 1.42        \\ \hline
\end{tabular}
\end{table}

Figure 5 shows the total energy difference between the FM and AFM-I magnetic phases with respect to the ground-state structure. The energy difference between these two phases is very small for each rotation angle, where the FM state is always the lowest in energy. The total energy difference between the ground state structure and this state is about 90 meV (22.5 meV per formula unit). The DOS observed for each rotation angle does not show significant changes from the results displayed in Fig. 4. Therefore, no electronic or magnetic phase transition is triggered via this rotation pattern.

Let us now consider the \textit{simultaneous rotation} case. In Table II we provide total energy differences between the FM and AFM-I states, relaxed cell parameters, and magnetization per Ru atom as a function of the RuO$_6$ octahedra rotation angles. We note that cell parameters highly depend on the rotation performed. Therefore, as we reduce the octahedra rotation in the system, the $c$ parameter decreases and $a$ and $b$ increase. We do not find a magnetic phase transition despite of the wide range of RuO$_6$ rotations explored. However, it is worth mentioning that the smallests energy differences between FM and AFM-I states are comparable in the accuracy threshold of $\simeq$ 1-2 meV. Therefore, different magnetic domains may coexist under any RuO$_6$ octahedra rotation performed in the system.

Figure 6a shows the DOS as a function of octahedra rotation angles. As we decrease the RuO$_6$ rotations the spin-up contributions spread out towards lower energies thereby closing the bandgap for the non-rotated structure. Figures 6b and 6c show the PDOS onto O atoms and Ru-$4d$ orbitals for ground-state and non-rotated structures respectively. The non-rotated structure, unlike the ground-state, has nearly regular octahedra with all Ru-O distances between 1.96-1.99 \r{A} and therefore, the crystal field splitting has almost degenerated t$_{2g}$ orbitals with e$_g$ (\textit{x$^2$-y$^2$}) levels slightly occupied at higher energies (see Fig. 6c). However, RuO$_6$ rotations introduce distortions in the octahedra producing larger splittings between e$_g$ and t$_{2g}$ levels that remove the e$_g$ electronic occupation (see Fig. 6b).

Based on these results, no magnetic phase transition is triggered when RuO$_6$ octahedra rotations increase/decrease with respect to the ground-state structure. We only observe small changes in the electronic properties represented by different metallic characters. Rotations about other axis (i.e., tilts) or a combination of distortions and rotations may be needed to drive the system towards a phase transition. One possible approach would be the application of uniaxial pressure along the $c$-axis, thus increasing (reducing) the out-of-plane (in-plane) polarization between Ru and O atoms which would favor the e$_g$ orbital population over the t$_{2g}$. However, more studies are needed to understand how distortions couple with the electronic and magnetic properties in Sr$_3$Ru$_2$O$_7$ as our recent work on octahedra tilts.\cite{our-surface}

\section{Conclusion}

We found that the ground-state structure of Sr$_3$Ru$_2$O$_7$ is characterized by a half-metallic character and a strong competition between FM and AFM magnetic phases. To examine the possibility of obtaining a magnetic or electronic phase transition we also investigated the role of the RuO$_6$ octahedra rotations as a function of the rotation angle. While the magnetic state does not depend on the octahedra rotations, the electronic properties change. We predict a half-metal--to--metal transition when the structure has no rotations due to the much higher crystal symmetry. These results indicate that rotations about the $c$-axis have small impact on the physical properties of Sr$_3$Ru$_2$O$_7$. However, other types of octahedra rotations (e.g. tilts) or/and distortions could lead the stoichiometric compound towards an AFM insulating phase. Our results motivate further studies on the coupling between octahedra distortions and properties in the Sr$_3$Ru$_2$O$_7$.

The computational work conducted by P. R and W. A. S. is supported by the U.S. Department of Energy under EPSCoR Grant No. DE-SC0012432 with additional support from the Louisiana Board of Regents and by an allocation of computing time provided by the Center for Computation and Technology located at the Louisiana State University. V. M. acknowledges support by New York State under NYSTAR program C080117. 

\medskip

\bibliographystyle{apsrev4-1} 
\bibliography{sample}

\end{document}